\documentclass[prl,aps,nofootinbib,twocolumn]{revtex4}
\usepackage{graphicx}
\usepackage{amsmath}

\begin{document}

\title{The pre-inflationary vacuum in the cosmic microwave background}

\author{Brian A. Powell} \email{bapowell@buffalo.edu}
\author{William H.\ Kinney} \email{whkinney@buffalo.edu}
\affiliation{Dept. of Physics, University at Buffalo,
        the State University of New York, Buffalo, NY 14260-1500}
\date{\today}
\begin{abstract}
\noindent
We consider the effects on the primordial power spectrum of a period of radiation-dominated expansion prior
to the inflationary era.  
If inflation lasts a total of only $60$ e-folds or so, the boundary condition for quantum modes
cannot be taken in the short-wavelength limit as in the standard perturbation calculation.
Instead, the boundary condition is set by the vacuum state of the prior radiation-dominated
epoch, which only corresponds to the inflationary vacuum state in the ultraviolet limit. This
altered vacuum state results in a modulation of the inflationary power spectrum. We calculate the
modification to a best-fit model from the WMAP3 data set, and find that power is suppressed at
large scales. The modified power spectrum is favored only very weakly by the WMAP3
temperature and polarization data. 
\end{abstract}
\maketitle

\section{Introduction}
The high degree of spatial flatness, isotropy and homogeneity that we observe in the universe today is well explained by inflationary cosmology \cite{Guth:1980zm}.  In order to obtain the universe that we observe, the
minimum number of e-foldings of inflationary expansion must lie somewhere between \(N = 46\) and \(N = 60\), but inflation could have lasted for significantly longer.  The nature of the universe prior to inflation is unknown, largely due to the fact that inflation is very good at
washing out initial conditions.  The existence of attractor solutions to the inflationary equations of motion make obtaining information about the pre-inflationary universe virtually impossible if inflation continues for long enough. Implementations of inflation in string theory ({\it e.g.}, Ref. \cite{Kachru:2003sx}) generically suffer from the ``$\eta$ problem'', which means they must be fine-tuned to achieve sufficient inflation. It is therefore well-motivated from the point of view of string theory to consider models with a minimal number of e-folds.  If inflation lasts just long enough to solve the horizon and flatness problems, it is quite possible that  pre-inflationary
cosmology will leave imprints on the present day universe.  In particular, pre-inflationary physics can lead to modulations in the primordial power spectrum, which may be observed in the large-scale
temperature anisotropies of the cosmic microwave background (CMB).

During inflation, cosmological perturbations had their start as tiny vacuum fluctuations which were then redshifted to super-horizon scales by the rapidly expanding spacetime.  Perturbations on the scale of the present-day
universe left the inflationary horizon at around \(N = 60\), having
originated as small-scale vacuum fluctuations at some earlier time.  In
order to fully reconstruct the primordial spectrum generated by
inflation, we must be able to understand the evolution of quantum modes
from their birth in the inflationary vacuum out to the extreme
large-scale limit. If inflation lasts for many more than 60 efolds, which
is commonly assumed, then the inflationary model fully specifies the
evolution on all observable scales.  However, if inflation lasts for just
the minimum number of e-foldings, then the initial conditions of the
quantum fluctuations must be set in the pre-inflationary era. Information
about this earlier time is therefore required to uniquely calculate the
inflationary power spectrum.

In this paper, we study the effects of a prior
period of radiation-dominated expansion on the subsequent inflationary epoch in models which yield the 
minimum amount of inflation, which we take to be \(N=60\) for definiteness.  
In such a
circumstance, the initial quantum modes are born not in the inflationary vacuum, but in the vacuum appropriate to radiation-dominated expansion.  
This has a profound effect on the largest-scale perturbations. The standard calculation of
perturbations in inflation assumes a well-defined short-wavelength limit for quantum modes.  However,
a fluctuation that leaves the horizon at the very start of inflation would need to have `started out'
large, since no non-inflationary mechanism could have brought it to such scales.  
Therefore, when we set initial conditions for the fluctuations at the start of inflation, we are forced to consider quantum fluctuations far from the short-wavelength limit.  As we show, fluctuations that
are born at large wavelength during the radiation-dominated era affect the power spectrum very differently
than fluctuations born at small wavelength during the inflationary era, resulting in a strong
suppression of power on large scales. 

The transition from a radiation-dominated phase to a
de Sitter universe was first studied by Vilenkin and Ford \cite{Vilenkin:1982wt}, Starobinsky \cite{Starobinsky:1982ee}, and by Linde \cite{Linde:1982uu}, and the production of gravity waves during such a transition has been studied by Sahni \cite{Sahni:1990tx}.  
These earlier works concentrated on the impact of thermal effects on the stability
of the de Sitter phase transition.  
In contrast, we consider the case of a decoupled scalar field and assume an instantaneous phase transition.  Not only does this simplify the analysis,
but it elucidates the profound effect that
the choice of vacuum alone has on the primordial power spectrum.  We emphasize the role of the choice of vacuum in the modulation of the power spectrum, as opposed to the pre-inflationary dynamics of the inflaton field itself, as considered by previous authors \cite{Contaldi:2003zv,Burgess:2002ub,Cline:2003ve}.
   
This paper is organized as follows: in Section I we review the standard treatment of inflationary
fluctuations in de Sitter space.  In Section II we study the evolution of both inflaton fluctuations
and curvature perturbations in the pre-inflationary radiation-dominated universe.  We find that if the inflaton is slowly
rolling then the two effectively decouple, allowing us to treat the inflaton as a massless
scalar during radiation domination.  This allows us to construct an adiabatic vacuum that is exact at
all wavelengths.  In Section III, we calculate the modification to the best-fit
WMAP3 power spectrum when we initialize the quantum modes in this vacuum.
We calculate the likelihood of the modified spectra for varying amounts of inflation and find a spread in
likelihoods \(|-2\Delta {\rm ln}\mathcal L| \lesssim 2\), the best-fitting
model yielding \(-2\Delta {\rm ln}\mathcal L = 1\). Despite the strong modification to the
underlying power spectrum, the $C^{TT}_\ell$- and $C^{EE}_\ell$-spectra are altered only
modestly, however, the \(C^{TE}_\ell\)-spectra are more sensitive to this
effect.  Therefore, power spectra with suppressed amplitude at large
scales are favored only weakly by current data, but improved
measurements of the polarization signal may be able to detect this effect
amidst the statistical scatter.  In Section IV we present conclusions. 
 
\section{Vacuum Selection in de Sitter Space}
In this section we briefly review the standard calculation of inflationary fluctuations,
in which the ultraviolet limit is assumed to exist for all quantum modes.

During inflation, quantum fluctuations in the field or fields driving
the expansion are quickly redshifted to scales far greater than the causal horizon, where they manifest themselves as classical curvature
perturbations.  Specializing to the case of a single scalar field \(\phi
= \phi_0 + \delta \varphi\),  the background field, \(\phi_0\), couples at linear
order to the metric perturbation.  It is therefore useful to introduce the gauge invariant Mukhanov potential,
\begin{equation}
\label{Muk}
u = a\delta\phi - \frac{{\phi'}}{H}\Psi,
\end{equation}  
where \(a\) is the scale factor of the universe, \(H\) is the Hubble
parameter and primes denote derivatives with respect to conformal time.  The field fluctuation, \(\delta \phi\), and potential,
\(\Psi\), are {\it gauge invariant} and defined as in Ref.
\cite{Mukhanov:1990me},
\begin{eqnarray}
\delta \phi &=& \delta \varphi + \phi_0'(B - E'), \\
\Psi &=& \psi - \left(\frac{a'}{a}\right)(B - E'),
\end{eqnarray}
where the perturbed metric is,
\begin{eqnarray}
ds^2 &=& a^2(\tau)\left[(1+2\phi)d\tau^2  \right. \nonumber \\
&-& \left. 2B_{,i}dx^id\tau-((1-2\psi)\delta_{ij}+2E_{,ij})dx^idx^j\right].
\end{eqnarray}
In comoving gauge, \(\delta \phi = 0\), and the metric perturbation measures the spatial curvature perturbation on
comoving hypersurfaces, \(\psi = \mathcal{R}\).  In this case we have
\begin{equation}
\mathcal{R} = \left|\frac{u}{z}\right|,
\end{equation}
where \(z = \phi'/H\).  The power spectrum of the curvature perturbation may then be written as a function of comoving wavenumber \(k\),
\begin{equation}
\label{scalspec}
P_{\mathcal{R}}(k) = \frac{k^3}{2\pi^2}\left<\mathcal{R}_{k}^2\right> = \frac{k^3}{2\pi^2}\left|\frac{u_k}{z}\right|^2,
\end{equation}
where the \(u_k\) are the Fourier modes of the gauge-invariant potential satisfying the equation
\begin{equation}
\label{modeq}
u_k'' + \left(k^2 - \frac{z''}{z}\right)u_k = 0.
\end{equation}

A non-zero correlation function \(\left<\mathcal{R}_{k}^2\right>\) means that there has been particle production due to the inflationary
expansion.  In an expanding
spacetime, the initial vacuum
state associated with positive frequency quantum modes will in general be associated with a mixture of
both
positive and negative frequency modes at later times.  An observer in a state initially devoid of
quanta will register a thermal bath of particles as the universe expands.  When solving Eq.
(\ref{modeq}) the initial conditions on the modes correspond to a specific choice of vacuum.     
As an example of vacuum selection, consider the simple case of a free scalar field fluctuation evolving in a de Sitter background, \(H = const\).  It is convenient to work
with the variable \(y = k/aH\), which is the ratio of the Hubble radius to the physical wavelength of the perturbation.  In a de Sitter
background the conformal time is 
\begin{equation}
\tau = -\frac{1}{aH},
\end{equation} 
so that \(y = -k\tau\) and Eq. (\ref{modeq}) becomes
\begin{equation}
y^2\frac{d^2u_k}{dy^2} + (y^2 - 2)u_k = 0.
\end{equation}
This has the general solution
\begin{equation}
\label{desittsol}
u_k(y) = \frac{1}{2}\sqrt{\frac{\pi y}{k}}\left[c_1H^{(1)}_{3/2}(y)+c_2H^{(2)}_{3/2}(y)\right]
\end{equation}
where \(H^{(1,2)}\) are Hankel functions of the first and second kind, respectively.  Since de Sitter
space is eternal, the initial conditions can be set in the infinite past, \(y \rightarrow \infty\).  In
this limit Eq. (\ref{desittsol}) becomes
\begin{equation}
\label{bd}
u_k(y) = \frac{1}{\sqrt{2k}}\left(c_1e^{iy} + c_2e^{-iy}\right).
\end{equation}
In order that the mode reduce to the vacuum state which annihilates positive frequency excitations, we must
choose \(c_2 = 0\), corresponding to the Bunch-Davies vacuum.  Then Eq. (\ref{desittsol}) becomes
\begin{equation}
\label{desittsol2}
u_k(y) = -\frac{1}{\sqrt{2k}}\left(1+\frac{i}{y}\right)e^{iy}.
\end{equation}
This solution starts off as a plane wave in the infinite past and at late times as \(y \rightarrow 0\), the second term in parentheses
comes to dominate and particle production occurs.

In general, not all exact solutions to
Eq. (\ref{modeq}) will asymptote to plane waves at initial times.  In this case, the best one can do is
construct an adiabatic vacuum to some finite order in a parameter which characterizes the slowness of
the expansion.  The adiabatic vacuum can be thought of as the vacuum that best approximates the
Minkowski vacuum in an expanding spacetime.  In order for such an approximation to be sensible, the
background expansion must be slow relative to the frequency of the quantum fluctuation.  From Eq.
(\ref{modeq}) with \(z''/z = C(\tau)\) and \(\omega = \sqrt{k^2 - C(\tau)}\), this condition can be
written
\begin{equation}
\label{adiabatic}
\frac{d {\rm ln}(C)}{d\tau} \ll \omega.
\end{equation}
The vacuum obtained above clearly satisfies this condition (in the infinite past the cosmological
expansion \(C(\tau) \rightarrow 0\)), and is equivalent to the lowest-order adiabatic vacuum.
In de Sitter space, the adiabatic condition Eq.
(\ref{adiabatic}) is \(-k\tau \gg 0\), which is not only satisfied in the infinite past, but at any later
time for infinitely-large momentum modes, \(k \rightarrow \infty\).  Therefore, as long as we take the
initial mode momenta sufficiently large, the adiabatic vacuum is a good approximation to the exact
solution.  This is to be expected, since at such small length scales the quantum modes do not feel the
expansion of the universe.  
However, modes that exit the horizon at the start of inflation will not have originated in the
short wavelength limit and are not well described by the state Eq. (\ref{bd}).  
The `initial state' of the quantum modes at the start of inflation must instead be
determined by the pre-inflationary physics.  In the next section, we consider the case
of radiation-dominated (RD) expansion prior to the inflationary epoch.

\section{Quantum fluctuations in a radiation dominated universe}
In a radiation dominated
universe, modes evolve under expansion from the long-wavelength limit \(y \ll 1\) to the
short-wavelength limit \(y \gg 1\).  Quantum fluctuations on the scale of the
inflationary horizon at the start of inflation
will therefore not have originated in the Minkowski vacuum in the ultraviolet.  We
nevertheless expect
quantum fluctuations to {\it exist} on all scales, including superhorizon, during the
radiation-dominated phase \cite{Mukhanov:1990me}.
Off-shell modes may be populated acausally as long
as they do not couple to classical perturbations on superhorizon scales.  These modes
will be later ``converted'' to classical perturbations during inflation.
In this section, we study the behavior of these fluctuations in the
RD phase, their evolution determining the initial state for the ensuing
inflationary expansion.  

The dynamics of a decoupled scalar field that is minimally coupled to gravity is
described by the action
\begin{equation}
S = \int \sqrt{-g} \,d^4x\left[\frac{1}{2}g^{\mu \nu}\partial_\mu \phi
\partial_\nu \phi - V(\phi)\right].
\end{equation}
Specializing to the case of a Robertson-Walker spacetime, variation with respect to the field \(\phi\) results in the Klein-Gordon
equation,
\begin{equation}
\label{kg}
\phi''+ 2aH\phi'+V_{,\phi}a^2 = 0,
\end{equation}
where primes denote derivatives with respect to conformal time.  The
equation of motion for the quantum fluctuations of the field are found by
perturbing the field about the homogeneous background solution, \(\phi =
\phi_0 + \delta \phi\), and linearizing Eq. (\ref{kg}) about this
background solution,
\begin{equation}
\label{flucs}
\delta \phi'' + 2aH\delta \phi'-\nabla^2\delta \phi +
V_{,\phi \phi}a^2\delta \phi -
4\phi'\Phi'+2V_{,\phi}a^2\Phi = 0.
\end{equation}
The background field couples at linear order to the gauge invariant metric perturbation,
\(\Phi\), which is defined as in Ref. \cite{Mukhanov:1990me},
\begin{equation}
\Phi = \phi + \frac{1}{a}\left[(B-E')a\right]'.
\end{equation}
We assume the absence of anisotropic stress, so that \(\Psi =\Phi\). 
Because of this coupling, we must also study the evolution of the metric
perturbation, \(\Phi\).  Furthermore, because the universe is radiation
dominated at this time, there will generically be fluctuations in the
radiation fluid as well. 
Perturbing Einstein's equations to first order, we obtain the equations of motion of
the metric perturbation,
\begin{eqnarray}
\label{ee}
&&\nabla^2 \Phi -3aH\Phi' - 3a^2H^2\Phi = \frac{4\pi}{m_{\rm Pl}^2}\left(a^2 \delta \rho_r
+ \delta \rho_\phi \right), \\
\label{ee2}
&&\Phi'' + 3aH\Phi' + 3a^2H^2\Phi + 2aH'\Phi = \frac{4\pi}{m_{\rm Pl}^2}\left(a^2 \delta p_r +
\delta p_\phi\right), \nonumber \\
\end{eqnarray}
where \(\delta p_r = \delta \rho_r/3\) denote the pressure and density
of the
radiation fluid, and for the scalar field we have 
\begin{eqnarray}
\delta \rho_\phi = -\phi'^2 \Phi + \phi'\delta \phi'+V_{,\phi}a^2\delta \phi, \\
\delta p_\phi = -\phi'^2 \Phi + \phi'\delta \phi'-V_{,\phi}a^2\delta \phi.
\end{eqnarray}

In this paper, we do not adopt a detailed model of pre-inflationary
physics and so the exact nature of the curvature perturbation is unknown.
However, one generically expects that both thermal fluctuations in
the radiation fluid as well as inflaton vacuum fluctuations will generate a
curvature perturbation across a range of scales.  
In what follows, we consider the case where the full range of comoving scales relevant
for CMB physics are initially superhorizon during radiation domination.  With \(k_{\ell = 2}/k_{\ell = 2000}
\sim 10^{-3}\) and if we suppose that \(k_{\ell = 2000}\) enters the horizon just below
the Planck scale, then \(k_{\ell = 2}\) enters the horizon at \(H \sim 10^{-6} m_{\rm
Pl}\).  Since we are considering a minimum amount of inflation, we require that the
quadrupole be on horizon scales at the start of inflation, giving \(H_{\rm inf} =
10^{-6} m_{\rm Pl}\).  This assumption simplifies the analysis somewhat, but relaxing
this condition does not significantly affect the qualitative results.

We first study the initialization of the thermal fluctuations.
The fractional thermal density perturbation in a comoving volume during
RD is given by
\cite{Peebles:1994xt},
\begin{equation}
\delta_T^2 = \left(\frac{\delta \rho_r}{\rho_r}\right)^2 = \frac{16}{3S},
\end{equation}
where \(S\) is the total entropy within the comoving region.  The entropy
density is given by 
\begin{equation}
s = \frac{2\pi^2}{45}g_{*s}T^3,
\end{equation}
where \(g_{*s}\) denotes the number of effective relativistic degrees of
freedom, and
the temperature \(T\) can be written,
\begin{equation}
T = \frac{1}{g_{*}^{1/4}}\sqrt{\frac{H m_{\rm Pl}}{1.66}}.
\end{equation}
Using these expressions with \(g_{*s} \approx g_{*}\), the initial amplitude of a thermal fluctuation
that evolves with the expansion to a physical scale \(k/a\) is
\begin{equation}
\delta_T \simeq \sqrt{160}y^{3/2}\left(\frac{H}{m_{\rm Pl}}\right)^{3/4},
\end{equation}
where \(y = k/aH\).  The physical scale of interest is that attained by the
fluctuation by the start of inflation, \(y_{\rm inf} = k/a_{\rm inf}H_{\rm inf}\).   
Furthermore, we know that fluctuations on the scale of the quadrupole
were on the scale of the horizon when inflation began, \(k_{\ell = 2} = a_{\rm
inf}H_{\rm inf}\).  This gives
\begin{equation}
\label{therm}
\delta_T \simeq \sqrt{160}\left(\frac{k}{k_{\ell = 2}}\right)^{3/2}\left(\frac{H_{\rm
inf}}{m_{\rm Pl}}\right)^{3/4}.
\end{equation}
It is important to note that although Eq. (\ref{therm}) is to be evaluated at the
beginning of inflation, it gives the {\it initial} amplitude of the thermal fluctuations on
each comoving scale.  Superhorizon modes should be initialized at horizon
crossing, since it is unlikely that the universe will have been in thermal equilibrium on
superhorizon scales prior to inflation.   

In order to determine the initial values of the inflaton fluctuations, one must make an
assumption about the pre-inflationary vacuum state.  If we suppose that the fluctuations
originate in the Minkowski `in-vacuum', a choice we will soon motivate, then they can be
canonically normalized, \(\delta \phi_i \propto (a_i\sqrt{2k})^{-1}\).  As mentioned, although
these fluctuations may exist on superhorizon scales, we do not expect them to couple to
the curvature acausally.  Like the thermal fluctuations, we therefore initialize them
at horizon crossing.

We are now ready to calculate the curvature perturbation generated by the
combined effect of the thermal and vacuum fluctuations. 
Consider the \(i\)-\(0\) perturbed Einstein equation,
\begin{equation}
\label{ij}
\left(\Phi' + aH\Phi\right)_{,i} = \frac{4\pi}{m_{\rm Pl}}\left(\delta u + \phi'\delta
\phi_{,i}\right),
\end{equation}
where \(\delta u\) denotes the velocity perturbation of the radiation fluid.  Since
thermal fluctuations are not vectorial, we can neglect the velocity
perturbation at early times. 
Substituting Eq. (\ref{ij}) into Eq. (\ref{ee}) and moving to Fourier
space yields the equation,
\begin{eqnarray}
\Phi_k &=& \frac{4\pi}{m_{\rm Pl}^2}\left(\frac{4\pi}{m_{\rm Pl}^2}\phi'^2 -
k^2\right)^{-1} \times \nonumber \\
&& \left[a^2 \delta \rho_{rk}+\phi'\delta \phi_k' + \left(V_{,\phi}a^2 +
3aH\phi'\right)\delta \phi_k\right].
\end{eqnarray}
The thermal and inflaton fluctuations enter this equation as
sources proportional to \(k^2\delta_{Tk}\) and \((V_{,\phi}a^2 + 3aH\phi')\delta \phi_k\),
respectively.  With \(\phi' \propto H_{\rm inf}\sqrt{\epsilon}\) and \(V_{,\phi}
\propto H^2_{\rm inf}\sqrt{\epsilon}\), the relative initial amplitudes of the sources may then be written,
\begin{eqnarray}
\label{rat1}
\frac{k^2}{(V_{,\phi}a^2 + 3aH\phi')}\frac{\delta_{Tk}}{\delta \phi_{k}}
&=& \frac{m_{\rm Pl}}{\sqrt{\epsilon}}\frac{H_i(k)}{H_{\rm inf}}\frac{\delta_{Tk}}{\delta
\phi_k} \nonumber \\
&=& \sqrt{\frac{320}{\epsilon}}\left(\frac{m_{\rm Pl}}{H_{\rm
inf}}\right)^{1/4}\left(\frac{k}{k_{\ell = 2}}\right)^{3/2}, \nonumber \\
\end{eqnarray}
where \(H_i(k)\) is the value of the Hubble parameter when mode \(k\) enters the
horizon.  We see that the initial thermal fluctuations easily dominate the initial vacuum
fluctuations.  For example, with \(H_{\rm inf} = 10^{-6} m_{\rm Pl}\) and \(\epsilon = 0.001\), the
above ratio \(\approx 2\times 10^{4} (k/k_{\ell = 2})^{3/2}\).  The
remaining terms involving \(\phi'\) and \(\phi'^2\), being proportional
to \(H_{\rm inf}^2\sqrt{\epsilon/k}\) and \(H_{\rm inf}^2\epsilon/k^2\), respectively, are
also suppressed.  We therefore expect the
inflaton fluctuations to decouple from the evolution of the pre-inflationary curvature
perturbation. 
In this circumstance, the curvature perturbation evolves as it would in a
single-component radiation universe, with the decaying solution \cite{Mukhanov:1990me},
\begin{eqnarray} \label{solphi}
\Phi({\bf x},\tau) &=& \tau^{-3}\left[\left(\frac{k\tau}{\sqrt{3}}{\rm
cos}\left(\frac{k\tau}{\sqrt{3}}\right) - {\rm
sin}\left(\frac{k\tau}{\sqrt{3}}\right)\right)C_1 \right. \nonumber \\
&+& \left. \left(\frac{k\tau}{\sqrt{3}}{\rm sin}\left(\frac{k\tau}{\sqrt{3}}\right) +
{\rm
cos}\left(\frac{k\tau}{\sqrt{3}}\right)\right)C_2\right]e^{i{\bf k}\cdot{\bf
x}}.\nonumber \\
\end{eqnarray}
Therefore, if the scalar and radiation perturbations are decoupled at horizon entry, they will remain decoupled as the modes evolve to subhorizon scales during the radiation-dominated evolution.

We are now in a position to determine the influence of \(\Phi\) on \(\delta \phi\).
We can gain insight into this question by comparing the magnitudes of the source terms in Eq. (\ref{flucs}).
Since \(\Phi\) is
decaying rapidly during RD, it suffices to compare the initial values of the source
terms.  Since the
thermal fluctuations are dominant, Eq. (\ref{ij}) reduces to the Poisson equation,
\begin{equation}
\label{poisson}
-k^2\Phi_k \approx \frac{4\pi}{m_{\rm Pl}^2}a^2\rho\delta_{Tk}.
\end{equation}
The ratio of source terms is then
\begin{equation}
\frac{2V_{,\phi}a^2_i\Phi_{ki}}{k^2\delta \phi_{ki}} \propto \sqrt{\epsilon}\left(\frac{H_{\rm
inf}}{H_i(k)}\right)^2\frac{\delta_{Tk}}{\delta \phi_{ki}},
\end{equation}
where we have neglected the term involving \(V_{,\phi \phi}\), which is small during
slow-roll.  By substituting \(\delta_{Tk}/\delta \phi_k\) from Eq. (\ref{rat1}), we find that this ratio can be easily made
\(\ll 1\). The term involving \(\Phi'\) in Eq. (\ref{flucs}) is
proportional \(H_{\rm
inf}\epsilon k\Phi_i\) and can be dropped, with the result that the \(\delta \phi_k\) are decoupled at linear order from the Newtonian potential \(\Phi_k\) created by the thermal perturbations in the radiation. The physical origin of this decoupling is easy to understand: during the period when the energy density of the universe is dominated by radiation, the Newtonian potential is likewise dominated by perturbations in the radiation field. However, in the limit that $\sqrt{\epsilon} \sim V'/V$ is small, the  scalar is to a good approximation decoupled from the Newtonian potential at linear order and evolves as a free field in the radiation-dominated background.  We show below that, despite the fact that the  potential is dominated by thermal fluctuations prior to inflation, the curvature perturbation after inflation is dominated by fluctuations in the scalar field. This is again true as long as $\epsilon$ is small. It is therefore 
 self-consistent to neglect thermal radiation perturbations in the analysis which follows, as long as the scalar field potential is sufficiently flat. 

With these simplifications, the equation of motion for the inflaton fluctuations Eq. (\ref{flucs}) becomes
\begin{equation}
\delta \phi_k'' + 2aH\delta \phi_k' + k^2\delta \phi_k + V_{,\phi \phi}a^2\delta \phi_k =0.
\end{equation}
Specializing to the case of radiation dominated expansion, \(a \propto
\tau\), and
redefining the field, \(u_k = a\delta \phi_k\), this becomes
\begin{equation}
u_k'' +\left(k^2 + a^2V_{,\phi \phi}\right)u_k = 0.
\end{equation}
As mentioned, it is safe to neglect the term proportional to \(V_{,\phi \phi}\) if the following
inequality is satisfied:
\begin{equation}
1 \gg 3\eta,
\end{equation}
where \(\eta \approx m^2_{\rm Pl}V_{,\phi \phi}/(8\pi V)\).
This is valid during RD as long as \(\eta\) does not vary significantly, a good
approximation during slow-roll.  If we then change our time variable, \(y = k\tau\), we
obtain the expression,
\begin{equation}
\frac{d^2u_k}{dy^2} + u_k = 0.
\end{equation}
This is the equation of motion of a massless scalar field in a radiation dominated background, and the main result of this
section.  This has the plane wave solution 
\begin{equation}
\label{pw}
u_k = c_1e^{-iy} + c_2e^{iy}.  
\end{equation}
This demonstrates the well-known fact that there is no particle production during radiation-dominated
expansion, since the quantum fluctuation never evolves out of its vacuum state. Since \(C(\tau) = 0\),
the adiabatic condition Eq. (\ref{adiabatic}) is satisfied for all time and for all \(k\).  
The vacuum is therefore well defined for all momenta and it remains to determine the
coefficients \(c_1\) and \(c_2\).  We know that at in the ultraviolet
limit, \(y =k\tau \rightarrow \infty\), the modes should reduce to the
Bunch-Davies form,
\begin{equation}
u_k(y) \underset{y \rightarrow \infty}{=}\frac{1}{\sqrt{2k}}e^{-iy}.
\end{equation}
However, it is not clear what the initial state of the large scale
fluctuations should be.  From the field decomposition,
\begin{equation}
\delta \phi({\bf x},\tau) = \int \frac{d^3k}{(2\pi)^{3/2}}\left(u_k(\tau)a_{\bf
k}+u^*_k(\tau)a^{\dagger}_{-\bf k}\right)e^{i{\bf k}\cdot{\bf x}}, 
\end{equation}
and the solution Eq. (\ref{pw}), we obtain the alternative decomposition,
\begin{equation}
\delta \phi({\bf x},\tau) = \int \frac{d^3k}{(2\pi)^{3/2}}\left(e^{-ik\tau}b_{\bf
k}+e^{ik\tau}b^{\dagger}_{-\bf k}\right)e^{i{\bf k}\cdot{\bf x}}.
\end{equation}
Here we have introduced the Bogoliubov rotated ladder operators,
\begin{eqnarray}
b_{\bf k} &=& c_1a_{\bf k} + c_2^*a^{\dagger}_{-{\bf k}}, \nonumber \\
b^\dagger_{-\bf k} &=& c_1^*a^{\dagger}_{-{\bf k}} + c_2a_{\bf k}.
\end{eqnarray}
A reasonable choice for an initial vacuum is \(b_{\bf k}|0 \rangle = 0\),
from which it follows that \(c_2 = 0\) for all \(k\).  This particular
choice of vacuum corresponds to the state of minimum uncertainty, and has
been used to study trans-Planckian effects
\cite{Danielsson:2002kx,Easther:2002xe}, as well as the quantum
decoherence of initial inhomogeneities \cite{Polarski:1995jg}.  Once we choose \(c_2
=0\), the value of \(c_1\) is determined from the commutation relation
between the field fluctuation and its conjugate momentum, 
\begin{equation}
[\delta \phi({\bf x},\tau),\pi({\bf x}',\tau')]_{\tau = \tau'} = i
\delta^3({\bf x} - {\bf x}'),
\end{equation}
where \(\pi = a^2 \delta \phi'\).  
With this choice of vacuum, we obtain the mode functions during RD,
\begin{equation}
\label{ics}
u_k(y) = \frac{1}{\sqrt{2k}}e^{-iy}.
\end{equation}
Other choices for the
initial state of the fluctuations are possible, and one ultimately must
make an assumption about the pre-inflationary physics.  This particular
choice is pure positive frequency on all scales, and asymptotes to the Bunch-Davies
limit in the ultraviolet. 

While we have determined that the curvature and inflaton
perturbations evolve independently during RD, we must not forget that the
pre-inflationary
curvature spectrum may contribute to the post-inflationary spectrum
generated by
inflation.  Since the pre-inflationary curvature perturbation decays (cf.
Eq.
(\ref{solphi})), we expect a negligible contribution on small scales.
However, on the
largest scales, the pre-inflationary curvature perturbation can be
approximated by
\begin{equation}
P_{\mathcal{R},{\rm pre}}(k) = \frac{k^3}{2\pi^2}|\Phi_{ki}|^2.
\end{equation}
The inflation-produced curvature perturbation spectrum is
\(P_{\mathcal{R}}(k) = H^2/(\pi \epsilon m_{\rm Pl}^2)\).
With \(\Phi_{ki}\) given by Eq. (\ref{poisson}) and considering scales
\(k \sim k_{\ell
= 2}\), we obtain the ratio,
\begin{equation}
\left. \frac{P_{\mathcal{R},{\rm pre}}(k)}{P_{\mathcal{R}}(k)}\right|_{k
\sim k_{\ell =
2}} \propto 10 \epsilon \left(\frac{m_{\rm Pl}}{H_{\rm
inf}}\right)^{1/2}.
\end{equation}
For \(H_{\rm inf} = 10^{-6} m_{\rm Pl}\), we find that \(\epsilon \ll
10^{-5}\)
if the spectrum produced by inflation is to dominate.

\section{The power spectrum and its effect on the cmb}
We are now in a position to determine the power spectrum that results from such a short period of
inflation.  The idea is to set the initial conditions, Eq. (\ref{ics}), for each mode at the start of
inflation, and then evolve them with Eq. (\ref{modeq}).  
As a concrete example, consider the exactly
soluble case of power law inflation, for which \(a \sim t^{1/\epsilon}\) and \(\epsilon = {\rm
const}\).  In this background, \(z''/z = a''/a\), and the general solution to Eq. (\ref{modeq}) is
\begin{equation}
\label{pli}
u_k(k\tau) = \sqrt{-k\tau}\left(c_1H^{(1)}_\nu(-k\tau) + c_2H^{(2)}_\nu(-k\tau)\right),
\end{equation}
with \(\nu = (3-\epsilon)/(2-\epsilon)\).  The coefficients are determined by matching to the
appropriate initial conditions, {\it i.e.} selecting a vacuum.
The power spectrum is obtained by evaluating Eq. (\ref{scalspec}) in the asymptotic limit, \(-k\tau
\rightarrow 0\).  In this limit, Eq. (\ref{pli}) becomes,
\begin{equation}
\label{asl}
u_k(k\tau) \underset{-k\tau \rightarrow \infty}{=} i\sqrt{-k\tau}Y_\nu(-k\tau)(c_1 - c_2),
\end{equation}
where \(Y_\nu\) is a Bessel function of the \(2^{nd}\) kind.  When \(c_1 = (1/2)(\sqrt{\pi/k})\) and
\(c_2 = 0\), we recover the standard result obtained when the quantum modes are assumed to have
originated in the Bunch-Davies vacuum.  We therefore rewrite Eq. (\ref{asl}) as
\begin{equation}
u_k(k\tau) \underset{-k\tau \rightarrow \infty}{=} 2\sqrt{\frac{k}{\pi}}u_{BD}(c_1 - c_2),
\end{equation}
where \(u_{BD}\) denotes the asymptotic form of the Bunch-Davies mode solution.  The power spectrum
follows:
\begin{equation}
P_{\mathcal R}(k) = 2\frac{k}{\pi}P_{\mathcal R}^{BD}(k)\left(|c_1|^2 + |c_2|^2 -c_1^*c_2 -
c_1c_2^*\right).
\end{equation}
This expression gives the modified power spectrum in terms of the standard spectrum, \(P_{\mathcal
R}^{BD}\), and a vacuum dependent term proportional to the coefficients.  This is valid for
{\it any} vacuum choice, with \(c_1\) and \(c_2\) generally being complicated functions of \(k\).  

\begin{figure}
\centerline{\includegraphics[angle=270,width=3.5in]{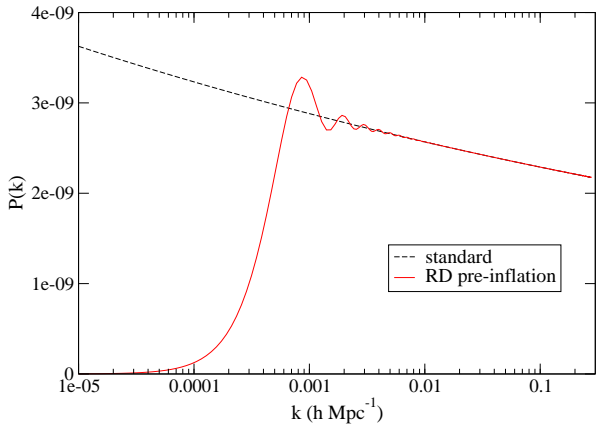}}
\caption{Best-fit WMAP3 spectrum calculated using standard inflationary initial conditions along with the same inflation model but with a
radiation-dominated pre-inflationary phase.}
\label{figure1}
\end{figure}

We now apply this approach to determine how the WMAP3 best-fit spectrum is modified if we
suppose that inflation was preceded by a radiation-dominated epoch.  We take the best-fit model with
scalar spectral index \(n_s = 0.951\), no tensors and no running \cite{Spergel:2006hy}.   
We assume that the best-fit spectrum is generated by some inflation model with sufficient inflation to
ensure that all observable scales originated in the Bunch-Davies limit of the inflationary vacuum.
This is equivalent to specifying the observable \(n_s\) at \(N = 60\). As
noted earlier, scales corresponding to the quadrupole can exit the
horizon at any time between \(N = 46\) and \(N = 60\).  Here we choose
\(N = 60\) but our results, Figures I-IV, are not sensitive to this
specific choice.  What is important is the amount of inflation occurring
{\it before} the quadrupole leaves the horizon.  We compare this result to
the case of the same best-fit inflationary model, but with a pre-inflationary RD phase ending at \(N
= 60\) (Figure \ref{figure1}). Instead of using the Bunch-Davies boundary condition for the modes, we
use the RD vacuum solution, Eq. (\ref{ics}), as a boundary condition.
The suppression of large-scale power is immediately evident in Figure \ref{figure1}.  The scales
corresponding to these fluctuations are vacuum modes that were in existence on scales of order the horizon
size at the onset of inflation.  As a result, their amplitudes are highly suppressed in relation to
quantum modes that would have attained these scales as a result of inflation.  Modes evolving during
inflation undergo mode freezing as they cross outside of the horizon, effectively locking-in their
amplitudes at relatively large values.  On sub-horizon scales, \(k > aH\), the modified spectrum
undergoes oscillations, rapidly approaching the standard inflationary spectrum.   While the
inflationary Bunch-Davies vacuum only exists for modes \(k \gg aH\), it has the same form as Eq.
(\ref{ics}).\footnote{There is an overall unobservable phase shift between the real
and imaginary parts of \(u_k\) owing to the fact that \(-k\tau
\rightarrow k\tau\) in a radiation dominated phase.} Therefore,  as \(k \rightarrow \infty\), the inflationary vacuum approaches the RD
vacuum, and the mode solutions become identical.  

A similar spectrum was obtained in \cite{Contaldi:2003zv}, in which the effects of an initially fast-rolling inflaton were studied.  Burgess {\it et al.} \cite{Burgess:2002ub} also obtain similar
spectra arising from a hybrid model in which a rapidly oscillating
auxiliary field leads to an era of matter domination prior to inflation.
We emphasize that we are encoding the effect of the pre-inflationary
physics solely in the choice of vacuum state for the inflaton
fluctuations. We  do not address the separate questions of how the
universe achieves the necessary level of homogeneity prior to the onset
of inflation, or of how the universe makes the transition from the
radiation-dominated to the inflationary phase (the former could be achieved via a previous period of inflation, as in ``open'' inflation models.) In this sense, our analysis should be regarded as a physically motivated ``toy'' model which demonstrates the effect of the pre-inflationary vacuum state on the primordial power spectrum.

Using a modified version of
\texttt{CAMB} \cite{Lewis:1999bs},  we are able to generate the \(C_\ell\)-spectra for prior RD.  In
Figure 2, we plot the temperature spectrum.    
\begin{figure}
\centerline{\includegraphics[angle=270,width=3.5in]{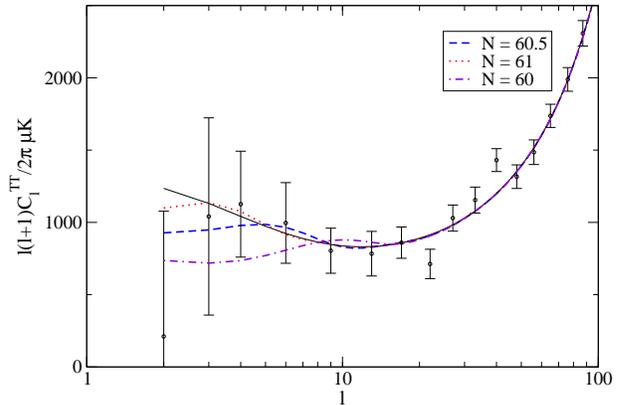}}
\caption{\(C^{TT}_\ell\)-spectra for the best-fit WMAP3 model (solid
black) together with models for which inflation is preceded by radiation
dominated expansion.  Models are plotted for varying durations of
inflation. 
Note the suppression of power at large scales.}
\label{figure2}
\end{figure} 
In each model considered in Figure 2, scales corresponding to the
quadrupole leave the horizon at \(N = 60\).  These models differ in the
total amount of inflation they provide, as indicated in the figure.   
The lack of power at large-scales is evident in the low-$\ell$
multipoles, but the spectra become virtually the same at around \(\ell
\sim 11\).  
The modification to the spectrum is strongly dependent on the duration of
inflation, and the effect is completely washed out for models yielding
more than 1 e-fold of inflation between the onset of inflation and the
time when scales corresponding to the quadrupole leave the horizon (\(N
\gtrsim 61\) in the figure.)  In Figures 3 and 4 we plot the
polarization auto-correlation (EE) and temperature-polarization
cross-correlation (TE) spectra.
Each model depicted in these figures lies within \(|-2\Delta {\rm ln}\mathcal
L| \lesssim 2\) of the best-fit model, and cannot be 
distinguished statistically with high confidence.  The temperature spectra all lie within the
cosmic variance envelope, making this effect difficult, if not
impossible, to decisively resolve with future experiments.  The EE
spectra are the least affected by the pre-inflation phase.  However, the TE cross-correlation spectrum
is more sensitive to this effect, exhibiting a stronger suppression at
low-\(\ell\). The limited
amount of available data makes it difficult to resolve these spectra at
the present time, however, it is plausible that improved polarization
measurements will be able to detect this effect in the future..
\begin{figure}
\centerline{\includegraphics[angle=270,width=3.5in]{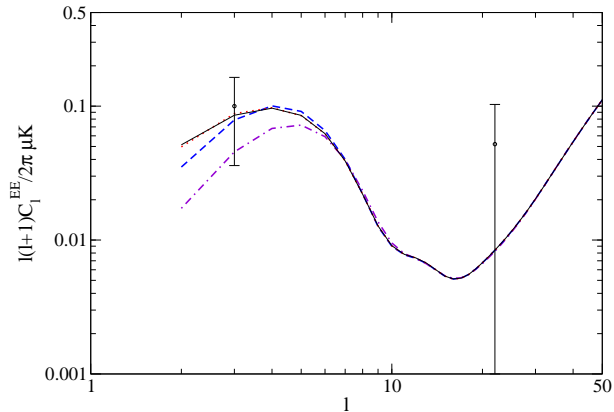}}
\caption{\(C^{EE}_\ell\)-spectra of the models presented in Figure 2.
The pre-inflation phase has a reduced effect on the EE spectra.} 
\end{figure}
\begin{figure}
\centerline{\includegraphics[angle=270,width=3.5in]{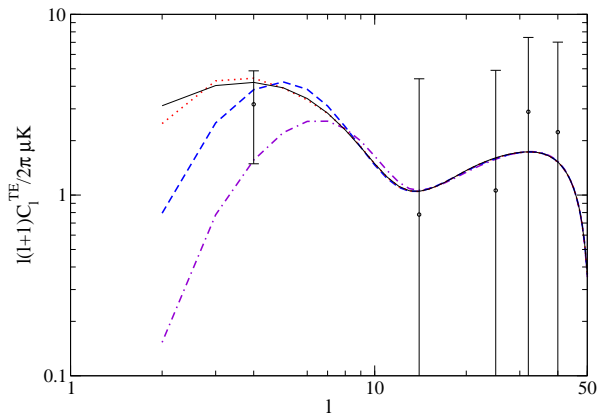}}
\caption{\(C^{TE}_\ell\)-spectra of the models presented in Figure 2.  The
suppression is more pronounced in these spectra than in the TT- and EE-spectra,
making this effect potentially observable with future data.} 
\end{figure}

This analysis can be easily extended to models with a tensor component.
Since tensor fluctuations are described by a massless scalar field, they
exactly satisfy the initial conditions Eq. (\ref{ics}).
Scalar and tensor modes on horizon and
superhorizon scales at the onset of inflation
experience the same degree of suppression, with the result that the
tensor-to-scalar ratio on these scales remains constant, \(r(k \lesssim
k_{\ell =2}) \approx
r(k_{\ell = 2})\).  This is in marked contrast to what is expected in other
pre-inflation scenarios.  Recently, Nicholson {\it et al.}
\cite{Nicholson:2007by} showed that an initially fast-rolling inflaton,
while yielding similar scalar spectra to those presented here, results in
a tensor-to-scalar ratio that increases in magnitude on large scales. Therefore, while improved E-mode polarization
measurements will be needed in order to detect pre-inflationary physics,
it appears that knowledge of the tensor component will be necessary for
determining what the pre-inflationary conditions were.  

\section{Conclusions}
\label{sec:conclusions}
We have considered the effects on the primordial power spectrum of a period of radiation-dominated expansion
prior to the inflationary epoch.  We impose initial conditions on the quantum fluctuations at the start of inflation appropriate to
the pre-inflationary era.  We find that if the inflaton is slowly rolling during RD then
it behaves as a massless scalar, decoupled from the pre-inflationary curvature
perturbations.  This
allows for the construction of an adiabatic vacuum that is exact at all wavelengths.  The largest-scale modes that exit the horizon at the
start of inflation are most strongly effected by the phase transition, since such modes were never in the ultraviolet limit.  This forces one to consider that the initially large-scale fluctuations
are vacuum fluctuations of very low momentum that originate in the adiabatic vacuum of the
radiation-dominated era.  These relatively low momentum modes lead to a strong suppression of
the power spectrum on the largest scales. We emphasize that our assumption of a decoupled scalar evolving in a background of thermal radiation fluctuations is self-consistent, but is certainly not the only possible set of conditions for the pre-inflationary phase. Absent a model for the pre-inflationary universe, one could in principle choose initial conditions with an arbitrary pre-inflationary curvature perturbation. This is true for any inflationary model: Trodden and Vachaspati, for example, have shown that the universe must be homogeneous on scales larger than the horizon size in the pre-inflationary phase in order for inflation to happen at all \cite{Vachaspati:1998dy}. 

The resulting power spectrum is shown to lower the \(C_\ell\)'s of the CMB temperature and
polarization anisotropies at low
\(\ell\), particularly around the quadrupole.  We find a spread in
likelihoods of \(|-2\Delta{\rm ln}{\mathcal L}| \lesssim 2\) in models
for which at most 1 e-fold of inflation occurs between the onset of
inflation and the time that scales corresponding to the quadrupole leave
the horizon, as compared to the best-fit WMAP model with
no tensors and no running.  We find an improvement of only \(-2\Delta{\rm
ln}{\mathcal L} = 1\) for models with \(N \approx 60.5\), so current data
does not strongly favor such models.  Since the modifications to the
\(C_\ell\)-spectra
occur at the cosmic variance dominated low-\(\ell\) multipoles, a more
accurate measurement of the more sensitive TE spectrum will be necessary
to decisively detect this effect. 

Hirai and Takami \cite{Hirai:2005pg} apply a methodology very similar to that presented here to investigate the effect of pre-inflationary physics, including that from a prior radiation-dominated phase. However, instead of considering a modification to the boundary condition of the quantum scalar mode $\delta\phi$, they use a boundary condition appropriate to a hydrodynamical radiation-dominated mode, with $c_s^2 = 1/3$. This results in the apparently unphysical conclusion that modulations to the inflationary power spectrum persist at all scales, regardless of the duration of inflation. However, we argue that the dominant contribution to the post-inflationary curvature perturbation will be from those generated by  fluctuations in the inflaton field itself, for which $c_s^2 = 1$ even in the pre-inflationary phase.

Finally, we draw attention to previous work on this subject,
namely \cite{Contaldi:2003zv} and \cite{Burgess:2002ub,Cline:2003ve}, who consider a pre-inflationary fast-rolling phase
and matter dominated phase, respectively. 
While the details of the pre-inflationary phase differ amongst these models, the effect on the primordial power
spectrum is strikingly similar.  However, Nicholson {\it et al.} \cite{Nicholson:2007by} showed recently that models
for which there is a transition
from an initial fast-rolling stage to the slow-roll attractor have a relatively large gravitational wave contribution
on large scales, in contrast to models preceded by radiation or matter-dominated expansion.
Our analysis differs further in that the suppression of power on large scales is manifestly the result of a choice of vacuum.  The details of the background dynamics do not enter into the calculation, except to specify the vacuum state in the pre-inflationary phase. The similarity of the modifications to the power spectrum suggests the conclusion that the effect of pre-inflationary physics can be generically considered as an effect of vacuum definition, in a fashion similar to to that for trans-Planckian modulations \cite{Easther:2002xe}. (This issue was considered in an effective field theory context in Ref. \cite{Kaloper:2003nv}.)
However, in all cases the resulting modification to the spectrum of fluctuations in the CMB results in only a modest improvement of the fit to the data. Since the observational uncertainties at these scales are dominated by cosmic variance, evidence for these effects is likely to remain inconclusive in the absence of additional observable evidence.

We thank Andrei Linde for helpful comments. This research is supported in part by National Science Foundation grant NSF-PHY-0456777.

\end{document}